# Improved Performance and Reliability of p-i-n Perovskite Solar Cells via Doped Metal Oxides.

By *Achilleas Savva[1], Ignasi Burgués-Ceballos[1] and Stelios A. Choulis[1]**

[1]A. Savva, [1]Ignasi Burgués-Ceballos, [1]S.A. Choulis, Prof.,

Molecular Electronics and Photonics Research Unit, Department of Mechanical Engineering and Materials Science and Engineering, Cyprus University of Technology, Limassol, 3603 (Cyprus).

E-mail: stelios.choulis@cut.ac.cy

**Abstract**

Perovskite photovoltaics (PVs) have attracted attention because of their excellent power conversion efficiency (PCE). Critical issues related to large area PV performance, reliability and lifetime need to be addressed. Here, we show that doped metal oxides can provide ideal electron selectivity, improved reliability and stability for perovskite PVs. We report p-i-n perovskite PVs with device areas ranging from 0.09cm$^2$ to 0.5cm$^2$ incorporating a thick aluminum doped zinc oxide (AZO) electron selective contact with hysteresis-free PCE of over 13% and high fill factor values in the range of 80%. AZO provides suitable energy levels for carrier selectivity, neutralizes the presence of pinholes and provides intimate interfaces. Devices using AZO exhibit an average PCE increase of over 20% compared with the devices without AZO and maintain the high PCE for the larger area devices reported. Furthermore, the device stability of p-i-n perovskite solar cells under the ISOS-D-1 is enhanced when AZO is used, and maintains 100% of the initial PCE for over 1000 hours of exposure when AZO/Au is used as the top electrode. Our results indicate the importance of doped metal oxides as carrier selective contacts to achieve reliable and high performance long lived large area perovskite solar cells.

## 1. Introduction

Organic-inorganic perovskite solar cells have recently emerged as a potential low-cost energy photovoltaic technology.[1] Impressive progress has been reported in just six years using solution-based production methods.[2] Power conversion efficiency (PCE) raised from 3.8% in 2009 [3] up to 21% in 2015.[4] This has sparked an enormous development in hybrid methylammonium lead halide perovskite materials $CH_3NH_3PbX_3$ (X = I, Cl, Br), revealing unique material properties such as tunable band gap[5], ambipolar nature[6] and charge carrier diffusion lengths larger than 1000 nm for mixed halides.[7-8]

These reports encouraged the intensive employment of the planar heterojunction (PHJ) architecture [9-10] which simplifies device fabrication process due to the removal of the mesoscopic layer.[11] Therefore, p-i-n structured device [12] and n-i-p structured device[13] can be fabricated depending on which polarity of contact is applied as substrate layer.

Typically, in both device structures, the perovskite absorber is sandwiched between two electrodes, each one selectively extracting one type of charge carrier. The n-i-p structure is usually based on ITO or FTO/compact $TiO_2$/perovskite/hole selective layer /Ag or Au.[14] In p-i-n structured perovskite solar cells the current flow is reversed by changing the polarity of the electrodes, being ITO or FTO/poly(3,4-ethylene dioxythiophene):poly(styrenesulfonate) (PEDOT:PSS)/ Perovskite/[6,6]-Phenyl-$C_{61}$-butyric acid methyl ester (PCBM)/Al the simplest device structure.[15]

In an entirely solution processed device, the choice and processing of charge selective contacts have a key role in device functionality. The bottom electrode charge selective layer requires highly transparent and conductive materials, which are processed on top of the transparent conductive oxide. In addition, the surface morphology and properties of the bottom selective layer must facilitate a compact and uniform perovskite film formation.[16]

On the other hand, the solution processing of selective layers on top of heat sensitive absorbers is a delicate task.[17-18] Moreover, an homogeneous coverage on the perovskite surface is necessary to minimize interfacial defects and to provide a good contact with the top metal electrode.[19] Another key parameter is the thickness of these buffer layers. For instance, Kim et al. proved that the thickness of hole selective layer in n-i-p structure has a major influence in device efficiency and reproducibility.[20]

The fullerene derivative PCBM is the most commonly used n-type material in the p-i-n structure, usually processed from a non-polar solvent solution on top of the perovskite layer.[21] It provides good contact with the perovskite layer facilitating the p-n junction at the interface, smoothening the surface and in many cases suppressing hysteresis.[22] On the other hand, it has been found that the PCBM thickness has a significant influence on device performance. Seo et al. demonstrated that increased PCBM thickness could result in FF and PCE drop derived from the limited electrical conductivity of the material.[23]

Several reports demonstrate that interface modification between PCBM and Al can enhance the top electrode functionality. Thermally evaporated Ca[24] and LiF[25] have been reported as interfacial layers in p-i-n perovskite-based solar cells. Concerning solution processed materials, Zhang et al proposed the use of different classes of conjugated polyelectrolytes as interlayers on top of PCBM to form an interfacial dipole which improved the selectivity of the top electrode.[26] In addition, Li et al proposed titanium isopropoxide TIPD as an electron transporting layer, which resulted in high FF p-i-n perovskite-based devices.[27]

Inorganic charge selective layers comprised of earth abundant materials have been reported in many cases to be beneficial in terms of device functionality. [28-29] In most of the cases metal oxides, which are commonly used in optoelectronic devices, suffer from limited electrical conductivity and thus only very thin layers can be used.[30] The limited electrical conductivity of the layers prohibits the vertical charge transport in thicknesses larger than ~50nm.[31]

One way to overcome the above limitations is by using doped metal oxides.[32-34] Due to their enhanced electrical conductivity, doped metal oxides can allow the fabrication of relatively thick conductive layers without significant increases in the device internal resistances. Highly transparent, more conductive and low-cost materials are more favorable for large area roll-to-roll (R2R) printing as well as device reproducibility.[35] In addition, one of the losses typically associated to upscaling of the device area is attributed to inhomogeneities in the active layer, including pinholes.[36] These negative effects can be potentially mitigated by introducing thick doped metal oxide layers, which additionally match product development targets.[37]

In this study we show that using a thick (over 150nm), low temperature processed aluminum-doped zinc oxide electron selective layer between PC[70]BM and Al can significantly impact the electrode functionality and device reproducibility. The proposed doped oxide exhibits exceptional conductivity and transmittance values when annealed at low temperature ($80^{o}C$ for 2 min), a critical parameter for solution processing on top of sensitive perovskite layers. A series of measurements including top view and cross sectional SEM, surface profilometry, J/V characteristics and statistical analysis reveal that using AZO as electron selective contact improves the functionality of the back electrode of p-i-n perovskite solar cells. Through detailed device physics analysis it is deduced that AZO provides an extra energetic barrier for positive charges according to the reduced leakage current derived from J/V measurements, enhances EQE in the near-infrared region by increased reflectance and light scattering effects as well as by smoothening the interface with the Al top contact. As a result devices using AZO as electron selective contact exhibit an average Jsc, FF and PCE increase of over 20% compared with the devices without AZO. Importantly, the smoother interface of AZO/Al compared to PC[70]BM/Al leads to a higher reproducibility of the results. Furthermore, when the device area is increased from 0.09cm$^2$ to 0.5cm$^2$, the PCE drops significantly in the PC[70]BM/Al based devices, while those incorporating the PC[70]BM/AZO/Al back electrode maintain a high performance. Furthermore, ISOS-D-1 lifetime data demonstrate enhanced stability of perovskite based solar cells when AZO is used as electron selective contact using different top metals. In particular p-i-n perovskite solar cells with AZO/Au top electrode show remarkable stable performance under ISOS-D-1 lifetime conditions.

2. **Results and discussion**

We compared two series of p-i-n mixed halide perovskite-based solar cells with the structure ITO/PEDOT:PSS/CH$_3$NH$_3$Pb(I$_{1-x}$Cl$_x$)$_3$/PC[70]BM/Al (PC[70]BM/Al-device) and ITO/PEDOT:PSS/CH$_3$NH$_3$Pb(I$_{1-x}$Cl$_x$)$_3$/PC[70]BM/AZO/Al (PC[70]BM/AZO/Al-device). **Figure 1** shows the energy levels of the materials comprising the solar cells under study as well as the optoelectronic properties of AZO layers prepared on quartz substrates.

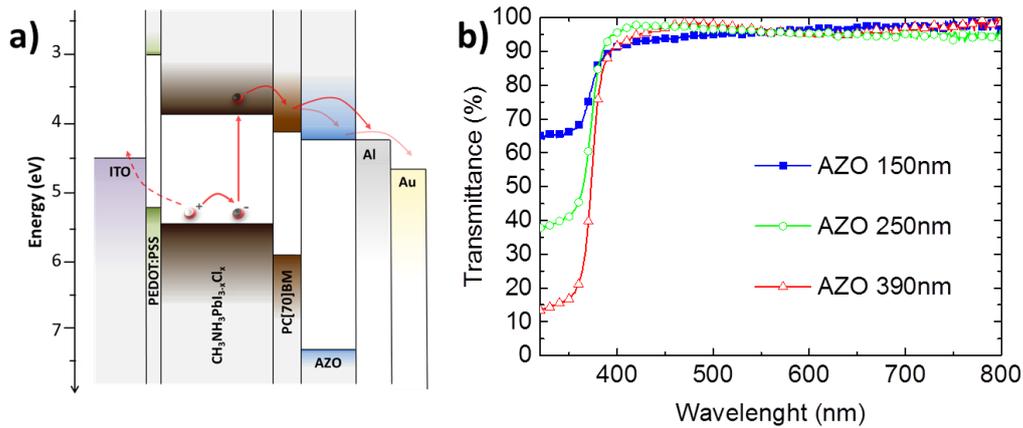

**Figure 1. a)** Band alignments of the solar cell. The data for $CH_3NH_3Pb(I_{1-x}Cl_x)_3$ and PC[70]BM, PEDOT:PSS and Al/Au are taken from ref [13], and for AZO are taken from ref [38]. **b)** Transmittance spectra of AZO layers with different thicknesses fabricated on quartz substrates, 150nm (blue filled squares), 250nm (green open circles) and 390nm layers (red open triangles).

Figure 1a schematically represents the hole and electron collection process within a functional perovkite solar cell device. The AZO layer has well matched energy levels (HOMO and LUMO values) for an efficient electron carrier selectivity (electron transporting/hole blocking capabilities). In comparison with PC[70]BM, the HOMO level of AZO is noticebly lower, yielding a stronger hole-blocking character to the buffer layer. Figure 1b shows the transmittance measurements performed on AZO layers prepared on quartz substrates using the same fabrication conditions as in our devices (see experimental section). As it is observed, the transmittance of AZO in the visible light spectrum (~400-800nm) is maintained over 90% even when the thickness of the layer is as high as 390nm. Furthermore, all three samples exhibit exceptional electrical conductivity values (~$1*10^{-3}$ S/cm) after only 2 minutes of annealing at 80°C as measured by the four point probe technique (Figure S1a). As mentioned before, selective layers with high electrical conductivity could provide the flexibility of thick layer incorporation without significant losses in FF and PCE. The high transparency of AZO allows to maintain high back electrode reflectivity, which can also contribute in the total photogenarated current of the device.[38] In general the presented optoelectronic properties of AZO thin film indicate potential benefits of using AZO thin films in perovskite solar cells, something which is in accordance with previous reports.[39]

Devices with and without a thick (~150nm) AZO electron selective contact were fabricated and tested. **Figure 2** demonstrates the average photovoltaic behavior of the two series of perovskite-based solar cells under study: ITO/PEDOT:PSS/CH$_3$NH$_3$Pb(I$_{1-x}$Cl$_x$)$_3$/PC[70]BM/Al (PC[70]BM/Al-device) and ITO/PEDOT:PSS/CH$_3$NH$_3$Pb(I$_{1-x}$Cl$_x$)$_3$/PC[70]BM/AZO(~150nm)/Al (PC[70]BM/ AZO/ Al-device). In this experimental run 12 devices for each series of devices were fabricated. The results were verified in more than 5 similarly executed experimental runs. Overall, over 60 devices with consistent results are reported within the manuscript.

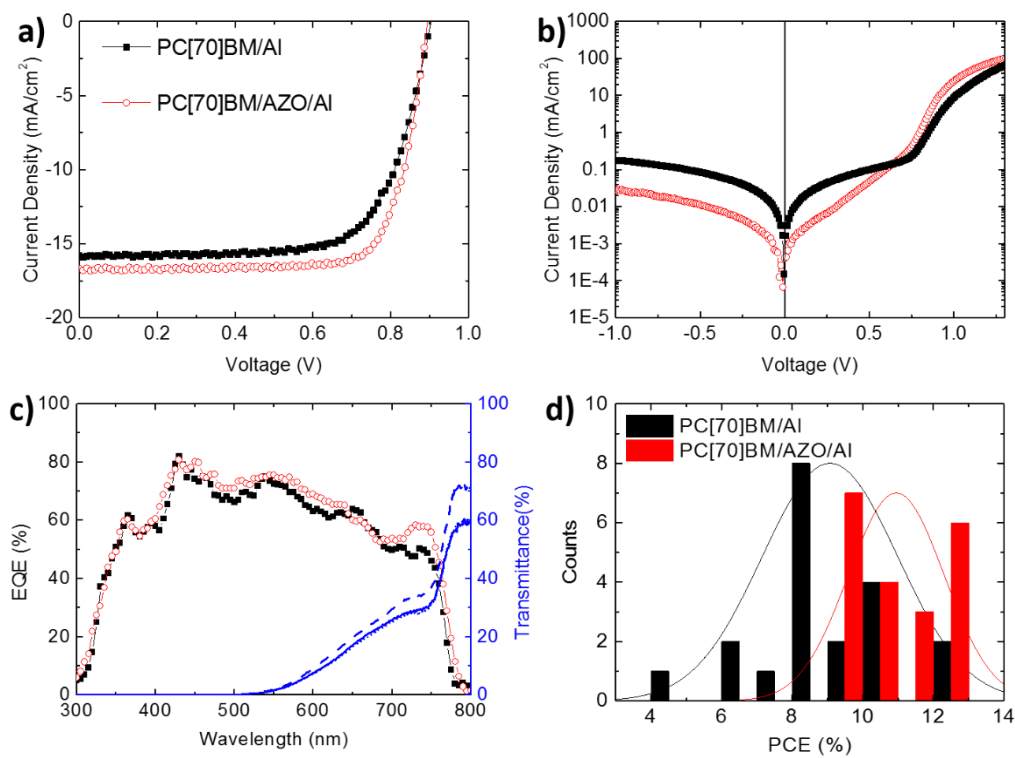

**Figure 2.** Current density versus voltage characteristics of the two compared devices ITO/PEDOT:PSS/CH$_3$NH$_3$Pb(I$_{1-x}$Cl$_x$)$_3$/PC[70]BM/Al (black filled squares) and ITO/PEDOT:PSS/CH$_3$NH$_3$Pb(I$_{1-x}$Cl$_x$)$_3$/PC[70]BM/AZO/Al (red open circles) **a)** under illumination and **b)** under dark conditions. **c)** External quantum efficiency (EQE) of the two solar cells under study and transmittance spectra of ITO/:PEDOT:PSS/CH$_3$NH$_3$Pb(I$_{1-x}$Cl$_x$)$_3$/PC[70]BM (solid line), ITO/PEDOT:PSS /CH$_3$NH$_3$Pb(I$_{1-x}$Cl$_x$)$_3$/PC[70]BM/Isopropanol washing (dot line), ITO/:PEDOT:PSS/ CH$_3$NH$_3$Pb(I$_{1-x}$Cl$_x$)$_3$/PC[70]BM/AZO (dashed line) and **d)** distribution of the PCE represented in histograms as obtained from 20 pixels (5 substrates) in two separate experimental runs.

**Figures 2a and 2b** demonstrate the J/V characteristics under illumination and dark conditions, respectively, for representative 0.09cm$^2$ devices with and without an AZO layer between PC[70]BM and Al. Both of the devices under comparison exhibited good diode behavior with high fill factor (FF), open circuit voltage (Voc) and short circuit current density (Jsc) values. The PC[70]BM/Al-based device exhibited Voc=0.9V, Jsc=15.81mA/cm$^2$, FF=70% and PCE=10.1%. On the other hand the PC[70]BM/AZO/Al-based device exhibited Voc=0.9V, Jsc=16.67mA/cm$^2$, FF=79% and PCE=11.9%. The increased FF is derived mainly from the decreased leakage current of the PC[70]BM/AZO/Al-based device compared to the PCBM/Al device under -1V bias as deduced from Figure 2b. The decreased leakage current (increased parallel resistance-Rp) within the device indicates better hole-blocking capabilities when incorporating the AZO electron selective layer within the top electrode. As predicted before, we ascribe this enhanced hole-blocking properties to the HOMO level of AZO, which is significantly lower than that of PC[70]BM. On the other hand, the series resistance (Rs) of the two devices under study were similar, as seen at +1V bias in Figure 2b. This indicates that the resistance to the vertical charge carrier flow from the perovskite layer to the terminals of the device remains constant in both type of devices. Despite the relatively high AZO thickness (>150nm), the vertical transport of electrons from the perovskite layer to the top Al contact is maintained due to the high conductivity values of the doped metal oxide (see Figure 1b). From these observations it can be deduced that the insertion of AZO between PC[70]BM and Al is enhancing the electron selectivity of the top electrode by providing better hole-blocking properties while high electron selectivity from the perovskite layer (where the majority of electrons are generated) to the top electrode is maintained. The combination of the increased Rp and unchanged Rs results in enhanced top electrode functionality and the best performing PC[70]BM/AZO/Al-based devices excibit Voc=0.9, Jsc=18.5 mA/cm$^2$, FF=80% and PCE=13.3% (Figure S3). Importantly, both type of devices under study exhibited negligible hysteresis (Figure S3). It was also observed that UV light does not have any influence on PCBM/AZO/Al-based devices, an effect which is usually related to n-type metal oxide electron transporting layers in both organic and perovskite-based solar cells.[40] All the average and champion photovoltaic parameters of the two p-i-n solar cells under comparison are summarized in table S1.

Figure 2c shows the External quantum efficiency (EQE) as a function of wavelength for the two type of devices. In accordance with figures 2a and 2b, the PC[70]BM/AZO/Al solar cell device exhibited higher EQE in almost all spectrum range compared to PC[70]BM/Al based devices The integrated Jsc values extracted from the EQE spectra of the two compared devices demonstrate a good match with the measured Jsc from the J/V characteristics, with a ~7% and ~6% deviation for the PC[70]BM/AZO/Al and the PC[70]BM/Al based devices, respectively.

The EQE shape is different in the region of 550-760nm in the case of PCBM/AZO/Al cells. To understand this different behavior we initially simulated the devices based on the optical properties of each layer using transfer matrix calculations.[42] The calculations do not reveal any change in the energy distribution within the device: the calculated Jsc is the same for a range of AZO thicknesses between 50-500nm (Figure S2). Similar results were published by Lin et al, where it was suggested that the thickness of the back electrode selective layer has a negligible impact in the distribution of the electric field within the device.[42] Transmittance measurements of the cell without Al back contact shown in figure 2c right axis also were also performed to understand the EQE shape difference in the region of 540-760nm. The transmittance of ITO/PEDOT:PSS/$CH_3NH_3Pb(I_{1-x}Cl_x)_3$/PCBM/AZO-based device structures is higher after ~600nm. On the other hand, the transmittance spectra of ITO/PEDOT:PSS/$CH_3NH_3Pb(I_{1-x}Cl_x)_3$/PCBM is identical when the device stack is washed with isopropanol, which is the solvent of AZO. The above observations suggest that the origin of the increased transmittance is not related to AZO solvent interaction with the device stack under-layers. The increased AZO transmittance is more likely due to AZO nanoparticles scattering effects. Thus the higher EQE observed in the region of 550-760nm AZO based cells can be attributed to additional absorption from the increased reflective light from Al electrode and scattered light from AZO nanoparticles within the interfaces of the back electrode.

Figure 2d shows the histograms for 20 devices contained in 5 different samples of the two compared series of solar cells produced in two separate experimental runs. Two main differences are therein observed. First, the device performance of PC[70]BM/AZO/Al devices is averagely higher by around 20% compared to those with PCBM/Al. Average increase has been observed in all experimental runs produced within the scope of this study. Second, the AZO containing devices show a remarkably narrower statistical data distribution. We observed systematically the same trend in several further experimental runs. These results confirm that the addition of a relatively thick (~150 nm) AZO layer between PC[70]BM and Al provides better device performance and, most relevantly, a substantial enhancement in device reliability. The latter is of an enormous importance in perovskite based photovoltaics, where low reproducibility is often reported.

These results clearly demonstrate the enhanced top electrode selectivity as well as a higher reproducibility of the inverted perovskite-based devices incorporating a relatively thick AZO layer between PC[70]BM and Al. As mentioned before, selective layers of such a thickness could lead to losses of FF due to increased Rs of the devices as a result of the limited electrical conductivity of the materials. Thus, since the material provides the flexibility of thicker layers to be incorporated without losses in device functionality, it was interesting to examine the effect of thickness of AZO electron transporting layer in device performance and reproducibility.

To proceed with our study, a range of PC[70]BM/AZO/Al-based devices with only varying the AZO thickness were fabricated. **Figure 3** demonstrates the J/V characteristics and the statistical distribution of the PCE represented in box plots for 3 different experimental runs with PC[70]BM/AZO/Al-based devices using different thicknesses of AZO.

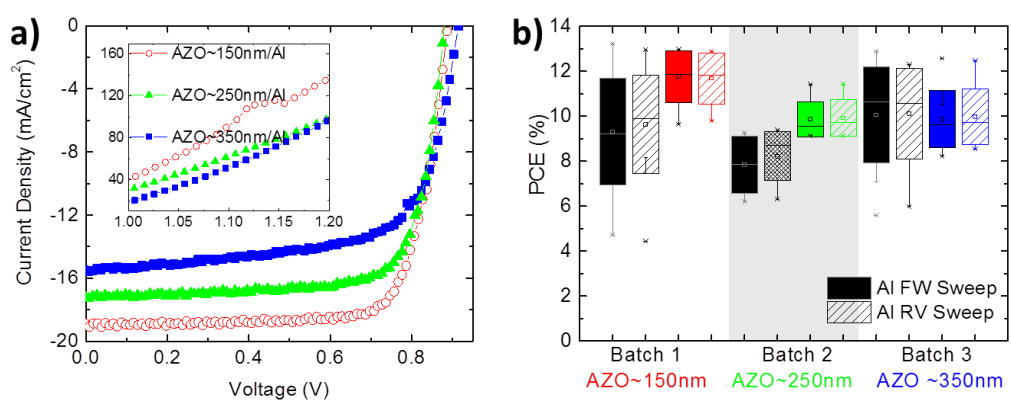

**Figure 3. a)** Current Density versus voltage characteristics of the solar cells with the structure ITO/PEDOT:PSS/$CH_3NH_3Pb(I_{1-x}Cl_x)_3$/PC[70]BM/AZO/Al with different AZO thickness: ~150 nm (red open circles), ~250 nm (green filled triangles), ~350 nm (blue filled squares). The inset represents the J/V characteristics of the same devices from +1 to +1.2V bias. **b)** PCE box plots of a set of data taken from 3 different experimental runs and 12 devices (3 substrates in each case) with different AZO thickness. Black boxes correspond to reference devices without the AZO layer.

**Figure 3a** demonstrates the representative J/V characteristics for three different AZO thicknesses. When the thickness of the AZO layer is increased from ~150 nm to ~350 nm, a simultaneous average reduction in short circuit current density and FF is observed. For the highest performing devices, the Jsc is reduced from 18.9

mA/cm$^2$ to 15.8 mA/cm$^2$, the FF from 80% to 57% and the PCE from 13.3% to 8.1%. The losses in device performance are mainly originated from an Rs increase in the devices using ~350 nm AZO compared with devices using ~150 nm of AZO as electron transporting layer as deduced from the inset in Figure 3a. This can be attributed to the increased resistance to the vertical current flow (Figure S1b). Similar effects have been reported in perovskite based solar cells for other buffer layers such as PC[70]BM [23] and Spiro-OMeTAD.[20] However, these materials only succeeded when thinner layers were employed, since they exhibit significantly lower electrical conductivity than AZO, as shown in Figure S1b. Consequently, the resistance starts limiting the device performance at much lower thicknesses, in the range of 50-100 nm. In contrast, we found that this happens in devices containing AZO only when layers thicker than 200 nm are used. Noteworthy, all the devices under study exhibited negligible hysteresis, as also demonstrated in Figure 3b. The box plots mean values for reverse and forward sweeps are almost the same in all the device types under study.

Despite lower performance was obtained when thicker AZO layers were used (Figure 3b), higher reproducibility was achieved in all AZO/Al cases compared to the reference PC[70]BM/Al based devices. The smoother interface of AZO/Al compared to PC[70]BM/Al could partially explain the increased device reproducibility. To verify this we examined the surface profile of all the layers comprising the perovskite-based devices. In addition, the back electrode layers surface topography and the cross sectional profile of the device were examined using SEM measurements. The data are demonstrated in **Figure 4.**

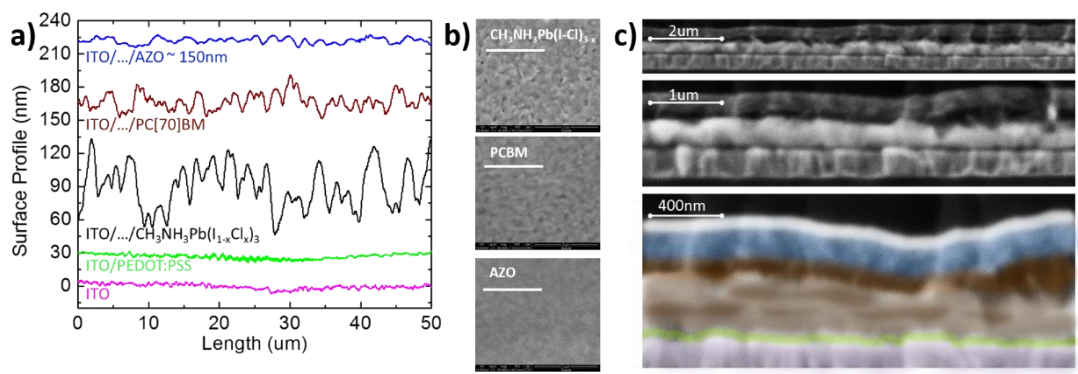

**Figure 4. a)** Surface profile of the layers comprising the perovskite-based devices under study. **b)** Top view of the layers comprising the top electrode of the devices under study (scale bar 3μm) and **c)** Cross–sectional SEM images of a complete solar cell ITO/PEDOT:PSS/CH$_3$NH$_3$Pb(I$_{1-x}$Cl$_x$)$_3$/PC[70]BM/AZO/Al from low (top) to high (bottom) magnification of the same device. The coloration assigned to each layer (online version) in the bottom image match with the colors in Fig 4a.

From Figure 4a it can be seen that ITO and PEDOT exhibited relatively smooth surfaces with an average surface roughness of about 10 and 5 nm respectively. In contrast, the perovskite layer exhibited high surface roughness above 50 nm. As it can be seen, the 50nm thick PC[70]BM layer smoothens significantly the perovskite layer and thus provides a relatively good interface with the top contact. Furthermore, the 150 nm AZO layer, which is deposited on top of the whole stack, further reduces the surface roughness down to ~15nm and provides an excellent contact with the top contact. The obvious spikes and the rough surface of PCBM is drastically smoothened by the thick AZO layer.

From the SEM images (Figure 4b and 4c) an estimation of the thickness of the layers within the device structure can be done (ITO ~250nm, PEDOT:PSS ~ 40nm, CH$_3$NH$_3$Pb(I$_{1-x}$Cl$_x$)$_3$ ~350nm PC[70]BM~50-80nm, AZO >150nm and Al ~100nm), which are in good agreement with the thickness measurements performed with the Dektak profilometer. In addition, the top view and cross sectional SEM images confirm that a thick and compact AZO layer was properly incorporated on top of PC[70]BM. It can be also observed that the well established AZO layer fills the visible interfacial defects and provides a more reproducible interface profile with Al compared to the PC[70]BM/Al interface.

All the data above suggest that AZO assists on the electron collection via important mechanisms. The thick AZO layer provides an extra energetic barrier for holes and thus improves the electron selectivity of the top electrode. In addition, the smoother interface of the AZO with the top Al contact significantly improves the reproducibility of the devices by reducing the interface recombination. Finally, the thick AZO layer avoids the direct contact between Al and perovskite spikes and pinholes, thus preventing shunting of the device.

Based on these observations, we were interested to examine the proposed top electrode in larger area devices. **Figure 5** shows the J/V characteristics and the photocurrent maps of large area devices (0.5 cm$^2$) with and without AZO between the PC[70]BM and Al. The different device areas were accurately defined by using custom made 3D printed masks (Figure S4).

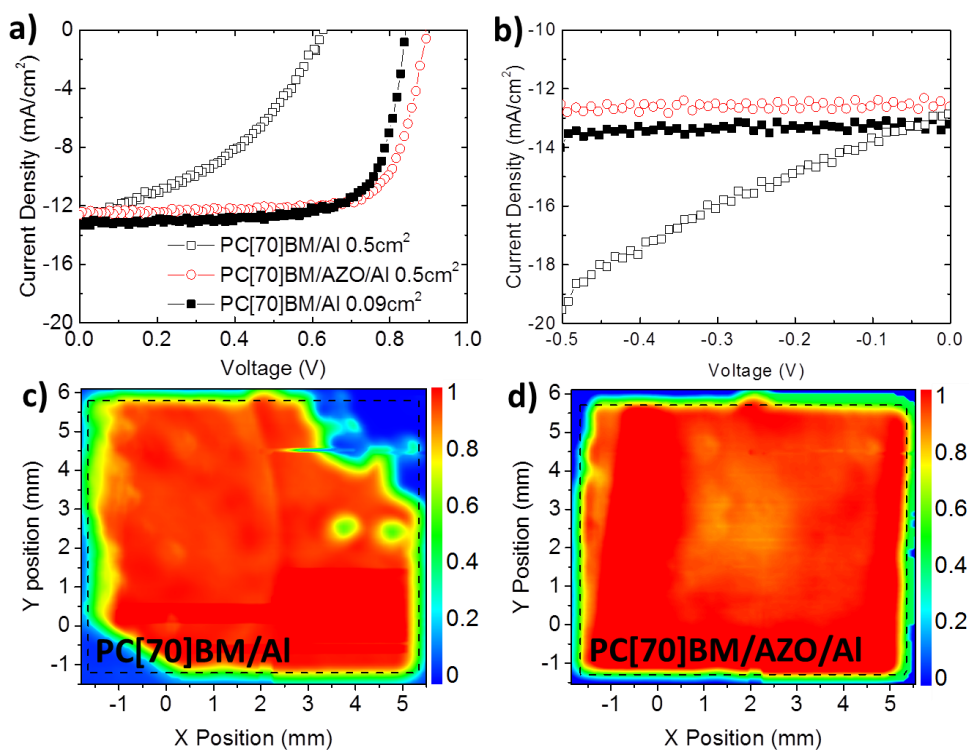

**Figure 5. a) and b)** Current Density versus voltage characteristics under illumination of a 0.5cm$^2$ device composed of ITO/PEDOT:PSS/CH$_3$NH$_3$Pb(I$_{1-x}$Cl$_x$)$_3$/PC[70]BM/Al (black open squares), 0.5cm$^2$ device comprised of ITO/PEDOT:PSS/CH$_3$NH$_3$Pb(I$_{1-x}$Cl$_x$)$_3$/PC[70]BM/AZO/Al (red open circles) and 0.09cm$^2$ device comprised of ITO/PEDOT:PSS/CH$_3$NH$_3$Pb(I$_{1-x}$Cl$_x$)$_3$/PC[70]BM/Al prepared in the same experimental run. **c)** Normalized photocurrent mapping of the 0.5 cm$^2$ PC[70]BM/Al and **d)** PC[70]BM/AZO/Al devices.

As shown in Figure 5a, a significant reduction of the performance in the 0.5 cm$^2$ PCBM/Al based devices was observed (Voc=0.63V, Jsc=13.2mA/cm$^2$, FF=42% and PCE=3.5%) compared with the representative 0.09cm$^2$ PC[70]BM/Al based device (Voc=0.85V, Jsc=13.5mA/cm$^2$, FF=74% and PCE=8.5%). The origin of the drop in Voc and FF in larger area (0.5cm$^2$) PC[70]BM/Al based devices is the increased leakage current (Rp) of the device compared with the small area (0.09cm$^2$)

PC[70]BM/Al based devices (Figure 5b). The increased leakage current could be attributed to the rough interface of PC[70]BM with Al (Figure 4) leading to an increased interface recombination and thus to a larger leakage current, lower FF, Voc and PCE. Although this effect is also visible in small area devices (figure 2b), it becomes a major limitation in larger area devices. The latter is verified by the photocurrent map (Figure 5c), where the formation of pinholes within the device area is clearly visible. These pinholes presumably arise from perovskite spikes which are not sufficiently covered by the PC[70]BM layer. These spikes are in direct contact with the Al top contact, minimizing the selectivity of the top electrode. On the other hand, the 0.5cm$^2$ PCBM/AZO/Al device maintained its performance, exhibiting Voc=0.9V, Jsc=12.9mA/cm$^2$, FF=78% and PCE=9.1%. In turn, the photocurrent map of the 0.5 cm$^2$ PC[70]BM/AZO/Al device (Figure 5d) reveals a more homogeneous distribution of photocurrent and absence of pinholes, in contrast to PC[70]BM/Al based device. We ascribe this improvement to the proved smoothened PC[70]BM surface by the thick AZO layer, which mitigates the negative effect of pinholes and ensures the selectivity of the top electrode. These results confirm the beneficial role of using a thick AZO layer also in large area devices.

Finally, the stability of devices with and without AZO were tested under the ISOS-D-1 protocol (shelf, T<25$^0$C, RH<50%). The devices were measured periodically for over 1000 hours of exposure. 8 devices from each series under study were examined and the average results are presented in **Figure 6**.

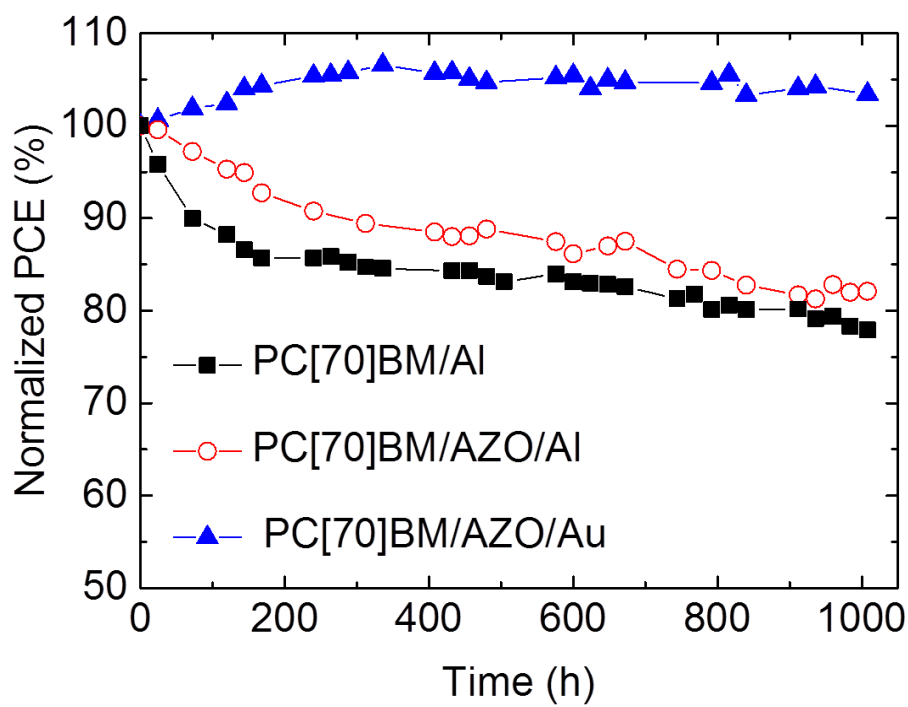

**Figure 6.** Average values of normalized PCE as a function of time of exposure under the ISOS-D-1 protocol for the encapsulated p-i-n perovskite-based cells, PC[70]BM/Al (black filled squares), PC[70]BM/AZO/Al (red open circles) and PC[70]BM/AZO/Au (blue filled triangles).

The PC[70]BM/Al-based device exhibits a significant drop in the normalized PCE within the first few hours. After 170 hours of exposure in air these devices have dropped down to ~85% of the initial PCE and only 78% of the initial PCE is maintained after 1000 hours of exposure. In contrast, PC[70]BM/AZO/Al-based devices maintain over 90% of their initial PCE after 170 hours of exposure. Though after 1000 hours of exposure PC[70]BM/AZO/Al-based devices also show significant degradation, performing only at 82% of their initial PCE.

As previously suggested Al metal is highly unstable and it's oxidation is a primary reason for the degradation of several optoelectronic devices.[43] In addition, reports suggesting interaction of the perovskite iodide with the Al metal is another degradation mechanism of perovskite based solar cells.[43] Since a thick AZO layer is in between the perovskite/PCBM layers and the Al we suggest that the delay in degradation of the p-i-n device when AZO is used could potentially be attributed to a delay in interaction between the perovskite and the top metal. Despite that, the Al oxidation is a separate mechanism and thus the devices using AZO also degrade after 1000 hours of exposure.

To further investigate the effect of AZO on device stability p-i-n devices with and without AZO using Au instead of Al as the top contact were fabricated. As shown in figure S5 devices without AZO are not functional when Au is used as the top contact and were excluded from the stability tests. Importantly, the incorporation of AZO allows the fabrication of efficient p-i-n perovskite solar cells with Au top contact (Voc =0.86V, Jsc= 16.2mA/cm2, FF= 75%, PCE=10.45% - figure S5). When AZO/Au is used as a top electrode the stability of the perovskite solar cells is significantly enhanced and after 1000 hours of lifetime testing under ISOS-D-1 protocol the p-i-n perovskite solar cells with AZO/Au top electrode maintain 100% of the initial PCE without losses (please see figure 6). In addition to these observations, a recent study by Bush *et al.* demonstrates that AZO can be used together with ITO top contacts for semi-transparent p-i-n perovskite solar cells with enhanced thermal stability.[34]

3. **Conclusions**

We have shown that aluminum-doped zinc oxide (AZO) is an ideal solution based electron selective contact for p-i-n planar perovskite-based PVs. Highly reliable p-i-n perovskite PVs using AZO electron selective contact with PCE of over 13 % and fill factor values approaching 80% are reported. The improved electrode functionality achieved by incorporating a thick (over 150 nm) AZO layer between PC[70]BM and Al results in an enhancement of the electron selectivity and device reproducibility. The proposed AZO electron selective layer exhibits exceptional transmittance and electrical conductivity values with only 2 minutes of annealing at 80$^{o}$C. This provides the flexibility to further increase the thickness of the electron selective layer over 150nm without losses in device functionality. The studies herein presented reveal that the proposed electron selective layer assists on the top electrode functionality by providing an extra energetic barrier for positive charges according to the reduced leakage current derived from J/V measurements, enhance EQE in the near-infrared region by increased reflectance and light scattering effects as well as by smoothening the interface with the Al top contact. These enhanced mechanisms, lead to enhanced Jsc and FF for PC[70]BM/AZO/Al-based devices and over 20% average increase in PCE compared to PCBM/Al based devices. Importantly, the effect of the thick AZO electron selective contact is very critical for high performance larger area devices (0.5 cm$^2$). The use of a thick AZO layer between PC[70]BM and Al ensures a homogeneous photocurrent generation and neutralizes the eventual, negative presence of pinholes. As a result, similar device performances compared with the small area devices are obtained. Finally, stability tests under the ISOS-D-1 protocol indicate that the incorporation of AZO in the back electrode of p-i-n perovskite-based solar cells improves lifetime performance. Importantly, AZO electron selective contact allows the use of more stable metals than Al, such us Au to be used as top electrodes and thus highly stable p-i-n based perovskite devices can be fabricated. We have reported p-i-n perovskite solar cells with AZO/Au top electrode with ultra-stable performance after 1000 hours of lifetime testing under ISOS-D-1 protocol.

To conclude, we have shown that doped metal oxides electron selective contacts can resolve major back electrode related barriers for p-i-n perovskite based solar cells. Doped metal oxides are found to lead in ideal electron selectivity, providing reliable large area highly efficient and stable perovskite-based solar cells.

## 4. Experimental Section

*Materials:* Pre-patterned glass-ITO substrates (sheet resistance 4Ω/sq) were purchased from Psiotec Ltd. The mixed halide perovskite precursor was purchased from Ossila Ltd. (product no I201). Aluminum-doped zinc oxide (AZO) ink was purchased from Nanograde (product no N-20X), PC[70]BM from Solenne BV and PEDOT:PSS from H.C. Stark (Clevios P VP Al 4083).

*Device Fabrication:* The whole stack of the inverted structure PV device was ITO/PEDOT:PSS/$CH_3NH_3Pb(I_{1-x}Cl_x)_3$/PC[70]BM/(AZO)/Al. We followed a device fabrication recipe provided by Ossila LTD. ITO substrates were sonicated in acetone and subsequently in isopropanol for 10 minutes and heated at 120$^o$C on a hot plate 10 minutes before use. To form a 40 nm hole transporting layer, the PEDOT:PSS ink was dynamically spin coated at 6000 rpm for 30s in air on the preheated ITO substrates and then transferred to a nitrogen filled glove box for a 25 minutes annealing at 120°C. The perovskite layer was then deposited on top PEDOT:PSS by dynamically spin coating the precursor solution at 4000 rpm for 30 seconds and annealing for 2h at 80$^o$C, resulting in a film with a thickness of ~350nm. The PC[70]BM solution, 50 mg/ml in chlorobenzene, was dynamically spin coated on the perovskite layer at 1000 rpm for 30s. In the devices containing AZO, the AZO ink (30 μl) was dynamically spin coated on top of PC[70]BM at 1000 rpm for 30s to fabricate ~50nm thin films. To increase the thickness of the AZO layer, several repetitions of the spin coating step were performed: layers of ~150nm, ~250nm and ~350nm were obtained by repeating the process 2, 4, and 6 times, respectively. All the AZO-based devices were annealed at 80°C for 2 minutes after performing all the coatings. Finally, 100 nm Al layers were thermally evaporated through a shadow mask to finalize the devices. Encapsulation was applied directly after evaporation in the glove box using a glass coverslip and an Ossila E131 encapsulation epoxy resin activated by 365 nm UV-irradiation. The active area of the devices was $0.09mm^2$ for the small and $0.5cm^2$ for the large scale devices. The AZO thickness for the large scale devices was ~150nm produced by 2 times repeated spin coating at 100rpm for 30sec.

*Characterization:* The thicknesses of the active layers were measured with a Veeco Dektak 150 profilometer. The current density-voltage (J/V) characteristics were characterized with a Botest LIV Functionality Test System. Both forward (short circuit → open circuit) and reverse (open circuit → short circuit) scans were measured with 10mV voltage steps and 40ms of delay time. For illumination, a calibrated Newport Solar simulator equipped with a Xe lamp was used, providing an AM1.5G spectrum at $100mW/cm^2$ as measured by a certified oriel 91150V calibration cell. A custom made shadow mask was attached to each device prior to measurements to accurately define the corresponding device area. Photocurrent mapping measurements were performed under 405nm laser excitation using a Botest PCT photocurrent system with a resolution of 40μm. Transmittance measurements were performed with a Schimadzu UV-2700 UV-Vis spectrophotometer. The electrical conductivity measurements were performed using a RM3000 four point probe system from Jandel Engineeering. Cross sectional topographic analysis was performed using a Quanta 200 Scanning Electron Microscope (SEM) (FEI, Hillsboro, Oregon, USA)

The specimens were mounted onto 45° aluminum stubs with an additional 45° tilt and imaged at 5kV beam voltage and 3.0 spot size.

**Acknowledgements**

This work has been funded by the H2020-ERC-2014-GoG project "Solution Processed Next Generation Photovoltaics (Sol-Pro)" number 647311. The SEM measurements performed within Professor A. Anayiotos BIOLISYS Laboratory experimental infrastructure facilities at Cyprus University of Technology. We would like to thank Efthymios Georgiou for his assistance on processing of perovskite formulation and Fedros Galatopoulos for his assistance on the stability measurements.

**Supporting information**

Supporting Information is available online from the Wiley Online Library or from the author.

**References**

[1]     G. Hodes, Science **2013,** 342, 317.

[2]     M. A. Green, K. Emery, Y. Hishikawa, W. Warta, E. D. Dunlop, *Prog. Photovoltaics* **2015,** 23, 805.

[3]     A. Kojima, K. Teshima, Y. Shirai, T. Miyasaka, *J. Am. Chem. Soc.* **2009,** 131, 6050.

[4]     EPFL, (Ed: D. Ltd), **2015.**

[5]     G. E. Eperon, S. D. Stranks, C. Menelaou, M. B. Johnston, L. M. Herz, H. J. Snaith, *Energy Environ. Sci.* **2014,** 7, 982.

[6]     F. Li, C. Ma, H. Wang, W. Hu, W. Yu, A. D. Sheikh, T. Wu, *Nat. Commun.* **2015,** 6.

[7] C. Wehrenfennig, G. E. Eperon, M. B. Johnston, H. J. Snaith, L. M. Herz, *Adv. Mater.* **2014,** 26, 1584.

[8] S. D. Stranks, G. E. Eperon, G. Grancini, C. Menelaou, M. J. P. Alcocer, T. Leijtens, L. M. Herz, A. Petrozza, H. J. Snaith, *Science* **2013,** 342, 341.

[9] D. Liu, T. L. Kelly, *Nat. Photon.* **2013,** 8, 133.

[10] K. Hwang, Y. S. Jung, Y. J. Heo, F. H. Scholes, S. E. Watkins, J. Subbiah, D. J. Jones, D. Y. Kim, D. Vak, *Adv. Mater.* **2015,** 27, 1241.

[11] X. Li, M. Ibrahim Dar, C. Yi, J. Luo, M. Tschumi, S. M. Zakeeruddin, M. K. Nazeeruddin, H. Han, M. Grätzel, *Nat. Chem.* **2015,** 7, 703.

[12] D. Forgacs, M. Sessolo, H. J. Bolink, *J. Mat. Chem. A* **2015,** 3, 14121.

[13] P. Docampo, J. M. Ball, M. Darwich, G. E. Eperon, H. J. Snaith, *Na.t Commun.* **2013,** 4.

[14] W. Zhang, M. Saliba, D. T. Moore, S. K. Pathak, M. T. Hörantner, T. Stergiopoulos, S. D. Stranks, G. E. Eperon, J. A. Alexander-Webber, A. Abate, A. Sadhanala, S. Yao, Y. Chen, R. H. Friend, L. A. Estroff, U. Wiesner, H. J. Snaith, *Nat. Commun.* **2015,** 6.

[15] L. Meng, J. You, T.-F. Guo, Y. Yang, *Accounts Chem. Res.* **2015.**

[16] M. Kaltenbrunner, G. Adam, E. D. Glowacki, M. Drack, R. Schwodiauer, L. Leonat, D. H. Apaydin, H. Groiss, M. C. Scharber, M. S. White, N. S. Sariciftci, S. Bauer, *Nat. Mater.* **2015,** 14, 1032.

[17] A. Savva, E. Georgiou, G. Papazoglou, A. Z. Chrusou, K. Kapnisis, S. A. Choulis, *Sol. Energ. Mater. Sol. C.* **2015,** 132, 507.

[18] A. Savva, I. Burgués-Ceballos, G. Papazoglou, S. A. Choulis, *ACS Appl. Mater. Interfaces* **2015,** 7, 24608.

[19] J. A. Christians, R. C. M. Fung, P. V. Kamat, *J. Am. Chem. Soc.* **2014,** 136, 758.

[20] G.-W. Kim, D. V. Shinde, T. Park, *RSC Advances* **2015,** 5, 99356.

[21] J. You, Z. Hong, Y. Yang, Q. Chen, M. Cai, T.-B. Song, C.-C. Chen, S. Lu, Y. Liu, H. Zhou, Y. Yang, *ACS Nano* **2014,** 8, 1674.

[22] J. Xu, A. Buin, A. H. Ip, W. Li, O. Voznyy, R. Comin, M. Yuan, S. Jeon, Z. Ning, J. J. McDowell, P. Kanjanaboos, J.-P. Sun, X. Lan, L. N. Quan, D. H. Kim, I. G. Hill, P. Maksymovych, E. H. Sargent, *Nat. Commun.* **2015,** 6.

[23] J. Seo, S. Park, Y. Chan Kim, N. J. Jeon, J. H. Noh, S. C. Yoon, S. I. Seok, *Energy Environ. Sci.* **2014,** 7, 2642.

[24] A. T. Barrows, A. J. Pearson, C. K. Kwak, A. D. F. Dunbar, A. R. Buckley, D. G. Lidzey, *Energy Environ. Sci.* **2014,** 7, 2944.

[25] X. Liu, M. Lei, Y. Zhou, B. Song, Y. Li, *Appl. Phys. Lett.* **2015,** 107, 063901.

[26] H. Zhang, H. Azimi, Y. Hou, T. Ameri, T. Przybilla, E. Spiecker, M. Kraft, U. Scherf, C. J. Brabec, *Chem. Mater.* **2014,** 26, 5190.

[27] C. Li, F. Wang, J. Xu, J. Yao, B. Zhang, C. Zhang, M. Xiao, S. Dai, Y. Li, Z. a. Tan, *Nanoscale* **2015,** 7, 9771.

[28] M.-H. Li, P.-S. Shen, K.-C. Wang, T.-F. Guo, P. Chen, *J. Mater. Chem. A* **2015,** 3, 9011.

[29] W. Qiu, M. Buffière, G. Brammertz, U. W. Paetzold, L. Froyen, P. Heremans, D. Cheyns, *Org. Electron.* 2015, 26, 30.

[30] H. Oh, J. Krantz, I. Litzov, T. Stubhan, L. Pinna, C. J. Brabec, *Sol. Energ. Mater. Sol. C.* 2011, 95, 2194.

[31] T. Stubhan, N. Li, N. A. Luechinger, S. C. Halim, G. J. Matt, C. J. Brabec, *Adv. Energ. Mater.* **2012,** 2, 1433.

[32]    A. Savva, S. A. Choulis, *Appl. Phys. Lett.* **2013,** 102, 233301.

[33]    T. Stubhan, I. Litzov, N. Li, M. Salinas, M. Steidl, G. Sauer, K. Forberich, G. J. Matt, M. Halik, C. J. Brabec, *J. Mater. Chem. A* **2013,** 1, 6004.

[34] K. A. Bush , C. D. Bailie , Y. Chen , A. R. Bowring , W. Wang , W. Ma , T. Leijtens , F. Moghadam, M. D. McGehee, *Adv. Mater.* **2016,** in press.

[35]    F. C. Krebs, *Sol. Energ. Mat. Sol. C.* **2009,** 93, 394.

[36]    Y. Galagan, E. W. C. Coenen, B. Zimmermann, L. H. Slooff, W. J. H. Verhees, S. C. Veenstra, J. M. Kroon, M. Jørgensen, F. C. Krebs, R. Andriessen, *Adv. Energ. Mater.* **2014,** 4, n/a.

[37]    W. Chen, Y. Wu, Y. Yue, J. Liu, W. Zhang, X. Yang, H. Chen, E. Bi, I. Ashraful, M. Grätzel, L. Han, *Science* **2015,** 350, 944.

[38] C. Waldauf, M. Morana, P. Denk, P. Schilinsky, K. Coakley S.A. Choulis, C.J. Brabec, *Appl. Phys. Lett.,* **2006,** *89,* 233517.

[39] X. Zhao, H. Shen, C. Zhou, S. Lin, X. Li, X. Zhao, X. Deng, J. Li, H. Lin, *Thin Solid Films* **2015,** in press.

[40]    S. Trost, K. Zilberberg, A. Behrendt, A. Polywka, P. Görrn, P. Reckers, J. Maibach, T. Mayer, T. Riedl, *Adv. Energ. Mater.* **2013,** 3, 1437.

[41]    G. F. Burkhard, E. T. Hoke, M. D. McGehee, *Adv. Mater.* **2010,** 22, 3293.

[42]    Q. Lin, A. Armin, R. C. R. Nagiri, P. L. Burn, P. Meredith, *Nat. Photon.* **2014,** 9, 106.

[43] V.M. Drakonakis, A. Savva, M. Kokonou, S.A. Choulis, Investigating Electrodes *Sol. Energy Mater. Sol. Cells* **2014,** 130, 544.

[44] Z. Jiang, X. Chen, X. Lin, X. Jia, J. Wang, L. Pan, S. Huang, F. Zhu, Z. Sun *Sol. Energy Mater. Sol. Cells* **2016,** 146, 35.